\documentclass{article}

\usepackage{PRIMEarxiv}

\usepackage[utf8]{inputenc} % allow utf-8 input
\usepackage[T1]{fontenc}    % use 8-bit T1 fonts
\usepackage{hyperref}       % hyperlinks
\usepackage{url}            % simple URL typesetting
\usepackage{booktabs}       % professional-quality tables
\usepackage{amsfonts}       % blackboard math symbols
\usepackage{nicefrac}       % compact symbols for 1/2, etc.
\usepackage{microtype}      % microtypography
\usepackage{lipsum}
\usepackage{fancyhdr}       % header
\usepackage{graphicx}       % graphics
\graphicspath{{media/}}     % organize your images and other figures under media/ folder

\title{Variation of the Atmospheric Boundary Layer Height at the Eastern Edge of the Tibetan Plateau
\thanks{\textit{Corresponding author}: 
\textbf{Fengrong Zhu. zhufr@home.swjtu.edu.cn}} 
}

\author{
  Jing Liu, Xiaofan Tang, Junji Xia, Fengrong Zhu*
 \\
  Affiliation \\
  Southwest Jiaotong University\\
  Chengdu 611756, Sichuan, China\\
  }

\begin{document}
\maketitle

\begin{abstract}
This paper utilized the high temporal and spatial resolution temperature profile data observed by the multi-channel microwave radiometer at the Large High Altitude Air Shower Observatory (LHAASO) on the eastern slope of the Tibetan Plateau from February to May and August to November 2021, combined with the ERA5 reanalysis data products for the whole year of 2021, to study the daily, monthly, and seasonal variations of the atmospheric boundary layer height (ABLH). The results are as follows: (1) The ABLH on sunny days showed obvious fluctuations with peaks and valleys. The ABLH continued to rise with the increase of surface temperature after sunrise and usually reached its maximum value in the afternoon around 18:00, then rapidly decreased until sunset. (2) The average ABLH in April was the highest at about 1200 m, while it was only around 600 m in November. The ABLH fluctuated greatly during the day and was stable at around 400 m at night. The ABLH results obtained from ERA5 were slightly smaller overall but had a consistent trend of change with the microwave radiometer. (3) The maximum ABLH appeared in spring, followed by summer and autumn, and winter had the lowest value, with all peaks reached around 14:00-15:00. These results are of great significance for understanding the ABLH on the eastern slope of the Tibetan Plateau, and provide reference for the absolute calibration of photon numbers of the LHAASO telescope and the atmospheric monitoring plan, as well as for evaluating the authenticity and accuracy of existing reanalysis datasets.
\end{abstract}

% keywords can be removed

\section{Introduction}
The Atmospheric Boundary Layer (ABL) is the lowest part of the atmosphere directly influenced by the Earth's surface. It serves as the immediate interface for surface-atmosphere interactions and facilitates the exchange of mass and energy between the Earth and the atmosphere \cite{1988An}. The Atmospheric Boundary Layer Height (ABLH) is an important parameter in atmospheric physics research, which includes applications in weather forecasting, pollution dispersion, and meteorological modeling.

To date, there is no instrument or method developed for direct measurement of ABLH. Instead, ABLH values are typically obtained indirectly through observations of atmospheric profiles, relevant parameters, or model calculations. The microwave radiometer is a passive ground-based microwave remote sensing instrument capable of receiving microwave signals radiated, scattered, or reflected from various heights. It provides vertical profiles of temperature, relative humidity, and water vapor density, and has become an important tool for ABLH retrieval. Cimini et al. \cite{2013Mixing}investigated the inversion of the convective boundary layer height using three instruments: a laser radar, microwave radiometer, and radio sounding. The results obtained from the microwave radiometer's observed brightness temperature data showed good consistency with the radio sounding results, surpassing the other two methods. Osibanjo et al. \cite{2022Intercomparison}utilized a combination of microwave radiometer, radio sounding, and micropulse lidar to retrieve the daytime convective boundary layer height and residual layer height in Mexico City. The results from all three instruments showed good correlation, with correlation coefficients exceeding 0.8. Furthermore, researchers have successfully used microwave radiometer temperature profiles to retrieve ABLH with reliable results.
As the plateau with the highest average elevation in the world, the Tibetan Plateau, also known as the "Third Pole," has a significant impact on atmospheric processes in the troposphere and lower stratosphere over the plateau region. It is also an important factor influencing climate change in China and globally. Chinese scientists have conducted extensive research on ABLH observations and analysis in the Tibetan Plateau region.

Chen et al. found that due to the higher elevation, drier climate, and stronger surface radiation, the ABLH in high-altitude plateau areas is significantly higher than that in plain areas and over the sea. ABLH in plain areas is typically around 1000-1500 m, while ABLH at Dangxiong (91.10°E, 30.47°N, 4300 m above sea level) can reach up to 2200 m. The complex terrain of the plateau enhances turbulent motion, and the strong turbulent activity contributes to the ABLH in plateau regions . The ABLH on the Tibetan Plateau ranges from 2000 to 3000 m, with notable differences between the dry and rainy seasons. This is because the surface sensible heat flux is greatly influenced by soil moisture. The higher the humidity, the lower the sensible heat flux, which suppresses boundary layer development, and vice versa\cite{2014Observations,2015Isolating}. Li et al. found that in August 2002, the ABLH at Nagqu (92.10°E, 31.47°N) reached a maximum of 1800 m. Using radio sounding data from the northern Tibetan Plateau, they further discovered that the ABLH during the dry season ranged from 2211 to 4430 m, while during the rainy season, it ranged from 1006 to 2212 m. Zhou et al.used occultation data to find that the large-scale circulation in the eastern Tibetan Plateau suppressed the development of the atmospheric boundary layer. Su Yanru et al. pointed out that cloud cover is the main factor affecting the ABLH and surface heat flux in the eastern Tibetan Plateau. Therefore, the height of the plateau boundary layer varies greatly under different climate conditions and locations, reflecting the different impacts of underlying surface conditions, solar radiation, cloud cover, large-scale circulation, and soil moisture on the structure of the boundary layer.

The eastern slope of the Tibetan Plateau is a transitional zone between the plateau and the Chengdu Plain and is an important component of the Tibetan Plateau. However, due to the area's high mountain ridges, sparse population, short meteorological observation history, and sparse observation stations, there are still shortcomings in the systematic study of the atmospheric boundary layer height in this region. Based on the high-altitude cosmic ray observation station 's multi-channel microwave radiometer temperature profile data with high temporal and spatial resolution observed from February to May and August to November 2021, and combined with ERA5 reanalysis data products for the entire year of 2021, the daily, monthly, and seasonal average daily variations of the ABLH were obtained in this paper. This provides a reference for better evaluating the authenticity and accuracy of the ABLH in reanalysis data and provides a basis for the absolute calibration of the telescope photon count of the high-altitude cosmic ray observation station and atmospheric monitoring plan.

\section{Data and Method}
\subsection{Observation Site}
The Large High Altitude Air Shower Observatory (LHAASO, 29.36° N, 100.14° E) is located at Haizi Mountain in the eastern part of the Tibetan Plateau. It is a forward-looking national major scientific and technological infrastructure during the 14th Five-Year Plan period. Utilizing its semi-array data, the LHAASO international collaboration group has discovered ultra-high-energy photons in the universe and ultra-high-energy accelerator sources in the Milky Way, opening a new era in ultra-high-energy gamma astronomy\cite{2021Ultrahigh}.

The MP-3000A microwave radiometer is installed inside the LHAASO. The LHAASO covers an area of over 2000 acres and is located on Haizi Mountain in the Sharuli Mountains, with an average altitude of 4410 m and a flat and vast terrain. The landform of LHAASO is mainly grassland, with shrubs, rocky areas, swamps, and river areas. Rivers flow through it, mainly distributed in the north, and the proportion of rocky areas, mountainous areas, and river areas is as high as 50$\%$. The climate conditions in this area are complex, with high soil moisture content, and precipitation concentrated from June to September. The average temperature in the hottest month is 8.6℃, and the average temperature in the coldest month is -7.1℃; the annual average highest temperature is 8.4℃, and the annual average lowest temperature is -3.9℃; the extreme highest temperature in a year is 20.9℃, and the lowest temperature is -22.9℃; the extreme highest ground temperature is 48.7℃, and the lowest is -30℃. The surrounding landform of LHAASO is shown in Figure 1, which is still dominated by plateau grassland, with deciduous and coniferous forests and cultivated land.

\begin{figure}
  \centering
  \includegraphics[width=0.5\textwidth]{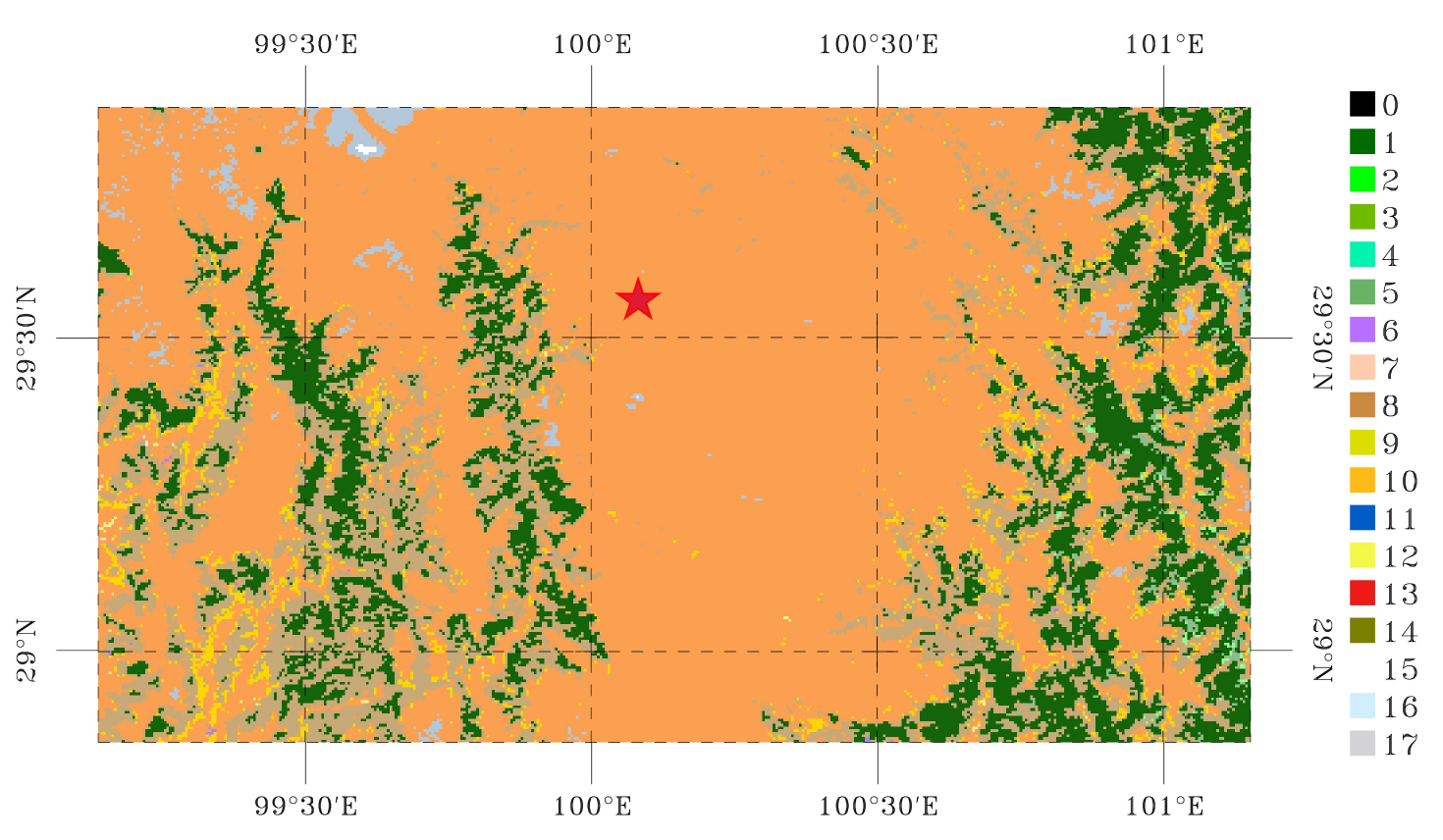}
  \caption{Geomorphological map of LHAASO observation station, the red five-pointed star marks the location of the LHAASO site, and the legend numbers represent different land cover types (0-water, 1-evergreen coniferous, 2-evergreen broad-leaved, 3-deciduous coniferous, 4-deciduous board, 5-mixed forest, 6-closed shrub, 7-open shrub, 8-savana, 9-savannah, 10-grassland, 11-wetland, 12-cultivated land, and so on). 13- Urban buildings, 14- Crops, 15- snow and ice, 16- poor or sparse, 17- unclassified)}
  \label{fig:fig1}
\end{figure}

\subsection{Data Sources}
\subsubsection{Microwave Radiometer Data}
The ground-based multi-channel microwave radiometer (MP-3000A 3105A) is a vertical remote sensing instrument for atmospheric temperature and humidity profiles, with 35 channels, including 21 K-band channels sensitive to water vapor and 14 V-band channels sensitive to temperature. It measures the atmospheric brightness temperature at 5 elevation angles (30°, 45°, 90°, 135°, 150°), and retrieves four kinds of profiles: atmospheric temperature, relative humidity, water vapor density, and liquid water density, as well as the total integrated water vapor and liquid water. 

The microwave radiometer provides observational data on temperature and humidity profiles within the troposphere below 10 km every approximately 2 minutes, with vertical resolutions of 0.05 km for the lowermost 0.5 km, 1 km for the range of 0.5-2 km, and 1.25 km for the range above 2 km. Overall, the spatiotemporal coverage of the microwave radiometer observations is better than that of radiosonde observations, and the spatial resolution of near-surface observations is higher than that of satellite observations. Due to instrument failures, maintenance, and other reasons, a large amount of data was missing in January, June, and July 2021, and the microwave radiometer has been out of operation since December 2021. Therefore, this study selected data samples from February to May and August to November 2021, which are currently only available to a small group of relevant researchers.

\subsubsection{ERA5 Reanalysis Data}
The European Centre for Medium-Range Weather Forecasts (ECMWF) ERA5 reanalysis data is a product that assimilates various observations including radiosonde data into a numerical weather prediction model, aiming to optimize the representation of the atmosphere and provide atmospheric data for times and regions without observations. This assimilation product achieves temporal and spatial coverage at a resolution of 0.25°×0.25° and a temporal resolution of 1 hour. To compare ERA5 data with microwave radiometer observations, the data from the four adjacent grid points to the microwave radiometer are extracted and the atmospheric boundary layer height (ABLH) at the observation site is calculated using a two-dimensional data interpolation algorithm based on area weighting.

\subsection{Microwave Radiometer Data Quality Control Methods}
Quality and reliability of microwave radiometer observations are affected by meteorological conditions, antenna shield maintenance, electromagnetic environment, and other factors. In this paper, a quality control method proposed by Fu et al. for MP-3000A microwave radiometer detection data is utilized to perform time and height consistency checks on the temperature observations, and to screen and eliminate suspicious profiles.
The time consistency check is used to screen out profiles with large temperature differences between adjacent time steps. The temperature variation of the microwave radiometer observation data with a time resolution of 2 minutes is analyzed in this paper. Within each month, the temperature variation between adjacent time steps should be within a certain range. If the absolute difference between the temperature difference between adjacent time steps and the mean temperature difference is greater than k times the standard deviation of the temperature difference, the data is defined as suspicious and the profiles that contain such data are determined as suspicious profiles.
\begin{equation}
    \left|dT_t-\overline{dT_t}\right|>k\times\sigma
\end{equation}
\begin{equation}
    dT_t=T_t-T_{t-1}
\end{equation}
\begin{equation}
    \sigma_1=\sqrt{\frac{1}{n}\sum_{t=1}^n (dT_t-\overline{dT_t})^2}
\end{equation}
In the equation, $T_t$ represents the temperature at time $t$ ; $dT$ is the temperature difference between adjacent time steps; $\overline{dT}$ represents the mean value of $dT$ ;$n$ is the sample size at a certain height for each month; $\sigma_1$ is the standard deviation of $dT_t$ ;$k$ is the multiple of deviation from the standard deviation, and in this paper, $k=3$ is taken.

Vertical consistency check is used to screen out the profile data with sudden temperature changes between adjacent levels. Temperature values should show a regular change with the increase of height, that is, the temperature difference between adjacent levels should not be too large. If the vertical change rate of temperature between adjacent levels is greater than its standard deviation, the data is defined as suspicious or erroneous data. The standard deviation is defined as shown in formula (4), and the vertical change rate of temperature is defined as shown in formula (5). The time consistency check is carried out layer by layer.
\begin{equation}
    \sigma_2=\sqrt{\frac{1}{m}\sum_{h=1}^m (dT_h-\overline{dT_h})^2}
\end{equation}
\begin{equation}
    dT_h=\frac{T_{h+1}-T_h}{H_{h+1}-H_h}
\end{equation}
where $T_h$ refers to the temperature at the $h$ level; $H_h$ refers to the height at the $h$ level;$\overline{dT}$ is the average of the temperature difference $dT$; $m$ is the sample size;$sigma_2$ is the standard deviation of $dT_h$.

\subsection{The Calculation Method for Atmospheric Boundary Layer Height}
ABLH stands for the maximum height that an upward thermal bubble can reach, or is defined as the height at which continuous turbulence stops. Different instruments reflect different aspects of the atmospheric boundary layer (ABL) properties, so the ABLH obtained by inversion may not be exactly consistent with each other numerically\cite{2012Lidar}. In this study, ABLH derived from the temperature profile obtained by microwave radiometer is used to reflect the intensity of atmospheric turbulence activity, which is correlated to some extent with PM2.5 and other particulate matter concentrations\cite{2016Background}, but cannot represent the distribution height of aerosols.

During the daytime, strong convection in the vertical direction is caused by the heat transfer from the warm surface, and turbulence in the ABL develops vigorously. Therefore, it is also called the convective boundary layer (CBL) with higher ABLH; at night, due to the ground radiation cooling, the atmosphere produces an inversion structure, and turbulence mixing is inhibited. The area extending from the surface to the top of the inversion layer is considered as the stable boundary layer (SBL) with lower ABLH. The part of the atmosphere above the stable boundary layer still retains the temperature characteristics of the daytime CBL, and is therefore called the residual layer (RL)\cite{DEARRUDAMOREIRA2018185}. The thermal conditions of the atmosphere are the most direct factors affecting ABLH. Since the flow states in the atmospheric boundary layer in different states are also different, the discussion on ABLH needs to be classified.

Several effective methods have been proposed by previous studies to calculate ABLH using the potential temperature gradient\cite{2010Observed,2020Analysis}. Therefore, in this study, the potential temperature gradient method is chosen to calculate ABLH. First, the stability of ABL is classified based on potential temperature gradient.
The stability of ABL is classified into three categories: convective boundary layer, stable boundary layer, and neutral boundary layer. According to the research results of Liu and Liang\cite{2010Observed}, the potential temperature gradient at two heights of 200 m and 50 m can be used to classify the stability of ABL according to the standard in equation (6), since LHAASO is a land surface.
\begin{equation}
    \theta_{200}-\theta_{50}<-\delta_s\quad(CBL)
\end{equation}
\begin{equation}
    \theta_{200}-\theta_{50}>+\delta_s\quad(SBL)
\end{equation}
\begin{equation}
    \theta_{200}-\theta_{50}=\quad else\quad(NBL)
\end{equation}
Selecting a height of 50 m is to remove the influence of uneven underlying surfaces; $\delta_s$ represents the potential temperature increment of the weak stable layer, and its value is taken as 0.25 K in this study.
The potential temperature $\theta$ can be obtained from temperature T and pressure P, and it is calculated using equation (7) \cite{1988An} in this study. The temperature data is obtained from measurements by microwave radiometers, and the pressure data can be calculated based on the atmospheric density (equation (8)). The atmospheric density data is obtained from the MSIS-E-90 model\cite{2021Atmospheric}.
\begin{equation}
    \theta_h=T_h\left(\frac{P_0}{P_h}\right)^{\frac{R}{C_P}}
\end{equation}
\begin{equation}
    P_h=g_h\int_{H_h}^{100}D_hdH_h
\end{equation}
Using professional terminology in the atmospheric field, translate the following Chinese text into English:
In the equation, $P_h$ is the pressure at height h, in units of Pa; $g_h$ is the gravitational acceleration at height h; $D_h$ is the atmospheric density at height h; $\theta_h$ is the potential temperature at height h, in units of K; $P_0$ is the standard pressure, taken as 1000 hPa; R is the gas constant; $C_P$ is the specific heat at constant pressure, with $R/C_P=0.286$. Then, the potential temperature gradient is calculated for each set of observation data to determine the stability of the atmospheric boundary layer.

For the CBL and NBL, following Liu and Liang\cite{2010Observed}: since buoyancy is the main mechanism driving CBL turbulence, we determine the ABLH as the height at which a neutrally buoyant air parcel rising adiabatically from the surface becomes unstable. We first search for the lowest layer h that satisfies Eq. (9) starting from the first layer (height 0 m), and then search upward again for the lowest layer that satisfies Eq. (10), which corrects for the h-layer.
\begin{equation}
    \theta_h-\theta_1\geq\delta_u
\end{equation}
\begin{equation}
    \theta'_h=\frac{\Delta(\theta_h)}{\Delta(H_h)}\geq\theta_r
\end{equation}
$\delta_u$ represents the minimum increase in potential temperature in the troposphere, with a value of 0.5 K adopted in this paper. $\theta_h'$ is the vertical gradient of potential temperature in the h-th layer. $\theta_r'$ is the minimum strength of the vertical gradient of potential temperature for the lid inversion, with a value of 4.0 K/km adopted in this paper. Here, $\theta'_s$ can be considered as the overshoot threshold of the rising air parcel. The ABLH for the SBL may be defined as the top of the overall stable layer starting from the ground or at the height of the maximum local wind speed, whichever is lower. However, due to the lack of wind speed data from the microwave radiometer, this paper defines the ABLH under the SBL as the former, assuming that buoyancy is the primary mechanism driving turbulence in the SBL. We first check whether $\theta'$ is greater than or equal to 0 for all levels, as shown in formula (11). If it is true, then the ABLH is the height at which $\theta'$ is minimized. If it is not true, then the ABLH is the height at which $\theta'$ equals 0 \cite{2020Analysis}.
\begin{equation}
   \theta'_h=\min(\theta')\quad or\quad0
\end{equation}

Based on the above quality control methods and the potential temperature gradient algorithm, this paper calculated potential temperature data and gradients using the microwave radiometer observation data from LHAASO between February 2021 and November 2021. The local ABLH variation sequence was diagnosed, and some patterns in the variation of the local ABLH were analyzed.

\section{Study on the Variations of Atmospheric Boundary Layer Height}
\subsection{Comparison Analysis of Monthly ABLH Variations Obtained from Microwave Radiometer and ERA Calculations}
Read the temperature profile data from the microwave radiometer and obtain daily hourly temperature profile data through two-dimensional linear interpolation. Then convert the hourly temperature profile data to monthly data for January, June, July, and December. Figure 2.1 shows the average monthly ABLH calculated based on the microwave radiometer observation data, represented by the red dotted line. It can be seen that the ABLH is significantly higher in spring (March to May) than in autumn and winter (September to November and February). The blue dotted line in Figure 2.1 shows the ABLH obtained from ERA5 reanalysis data, which has a similar trend to the results from the microwave radiometer, but is generally lower in height.

\begin{figure}
  \centering
  \includegraphics[width=0.5\textwidth]{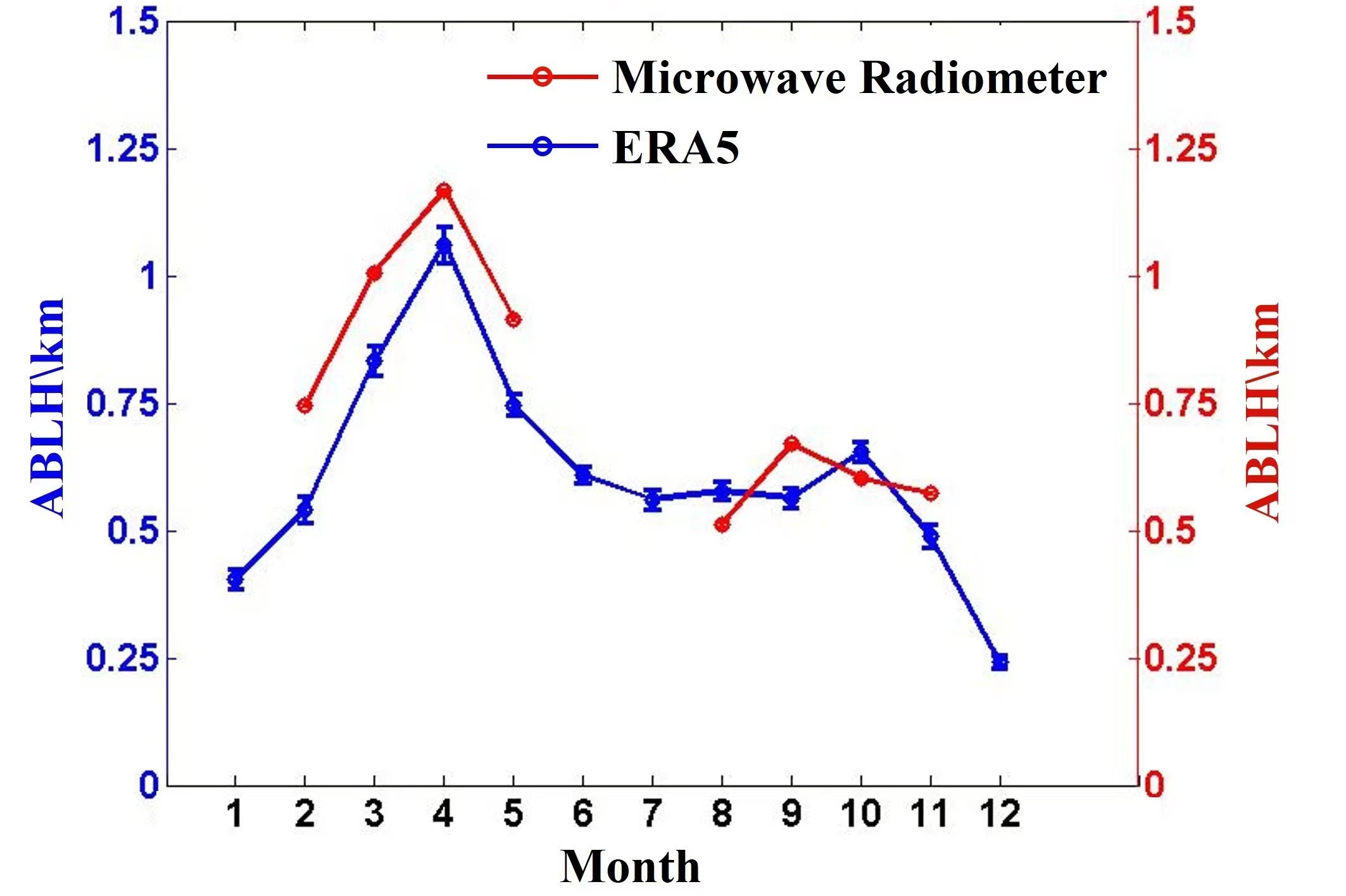}
  \caption{2021 Monthly average ABLH}
  \label{fig:fig2}
\end{figure}

\subsection{Comparison of Diurnal and Nocturnal ABLH Monthly Variations}

\begin{figure}
  \centering
  \includegraphics[width=0.45\textwidth]{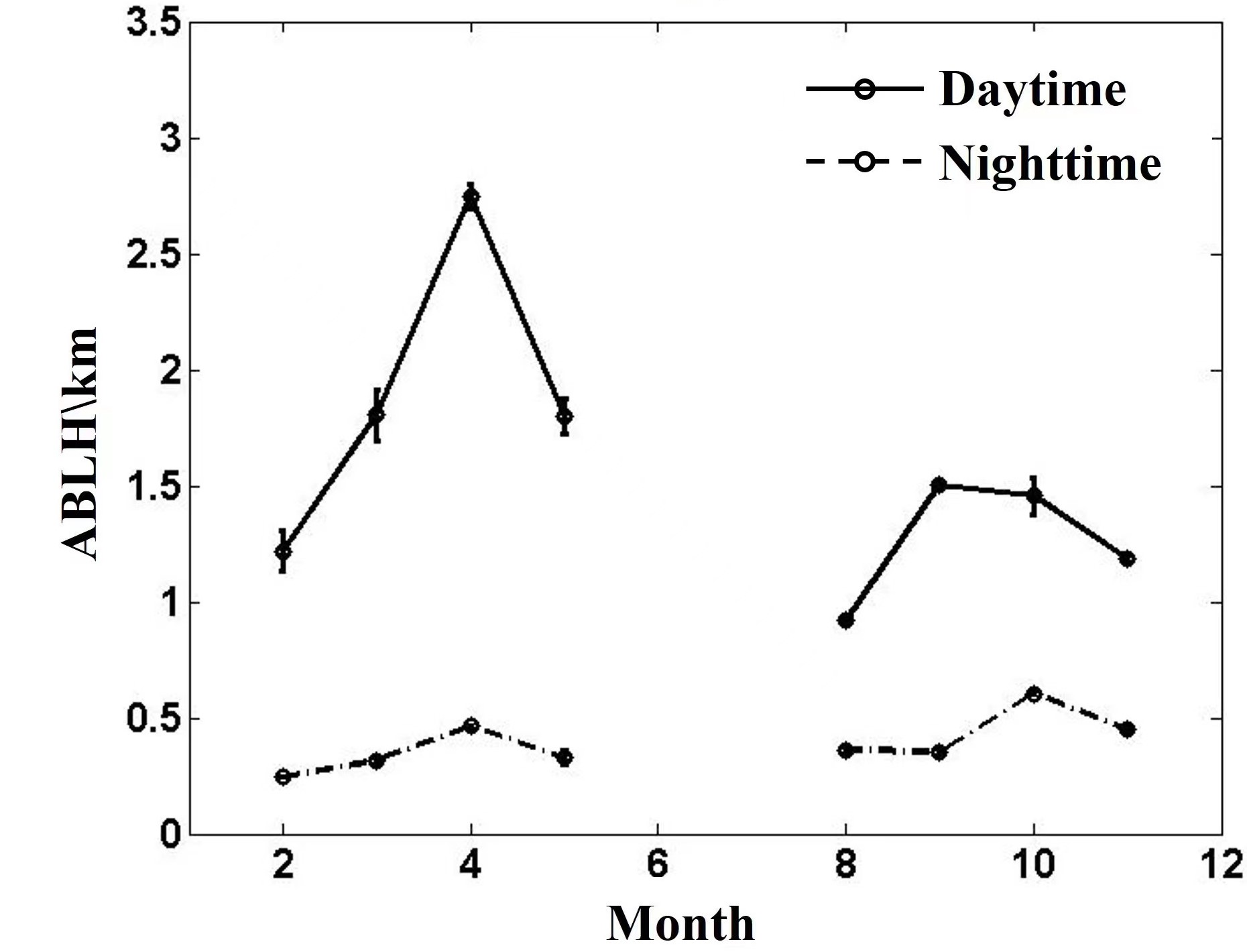}
  \caption{Monthly mean ABLH for day and night in 2021 from microwave radiometer data}
  \label{fig:fig3}
\end{figure}

In order to observe the monthly average diurnal and nocturnal differences of ABLH, the data was divided into two parts based on the sunrise and sunset times of the LHAASO observation station, and the average ABLH was calculated separately. Figure 3 shows the monthly average variations of ABLH during the day and at night. It can be seen that the ABLH during the day is significantly higher than at night in all months, and the ABLH at night is relatively stable at around 400 m, while the fluctuation amplitude of ABLH during the day is more intense than at night. The highest monthly average difference between diurnal and nocturnal ABLH appears in April, reaching more than 2 km, while the lowest difference appears in August, less than 0.5 km. 

The reasons for the above-mentioned trends may be twofold. On the one hand, solar radiation plays a role. In spring, solar radiation increases, sunrise time is early, the surface is heated by strong solar radiation for a long time, which is conducive to convection development and leads to higher ABLH. In autumn and winter, solar radiation weakens, surface temperature is low, convection activity weakens, and ABLH decreases. On the other hand, soil moisture also affects the evolution of the boundary layer by changing the energy balance. When soil moisture is low, the surface soil evaporation is also low, and more sensible heat flux is transmitted, which is conducive to the development of the boundary layer. When soil moisture is high, the latent heat is larger and the sensible heat is smaller, which suppresses the development of the boundary layer. The LHAASO station is mainly influenced by the westerly south wind in April-May and the plateau summer monsoon wind after that. The westerly south wind is faster and drier. Zhang et al.proposed that with the increase of wind speed\cite{Zhang2013Trends}, relative humidity and surface temperature, the boundary layer height increases; and Fuwei et al. proposed that the sensible heat flux is greater and ABLH is higher under the westerly south wind. The plateau summer monsoon wind is more humid, and the regions under this wind direction have more rainfall, soil moisture increases, the latent heat flux is larger, and the development of the atmospheric boundary layer is suppressed, leading to lower ABLH. Therefore, after reaching the peak in April, ABLH shows a continuous downward trend until August, with a slight increase in September, and then continues to decrease. From Figure 2, it can be seen that ERA5's ABLH shows a slight increase in October, which is different from the observation results of the microwave radiometer.

\subsection{Seasonal Average Daily Variation Analysis of ABLH}
Figure 4 shows the average daily variation of ABLH in different seasons, where blue represents the ERA5 reanalysis data and red represents the results calculated from microwave radiometer data. The dashed lines at the top and bottom of the graph represent the maximum and minimum values of ABLH at the same time each day during the season. The upper and lower ends of the box represent the corresponding upper and lower quartile values, and the horizontal line in the box represents the corresponding median value. The dotted line represents the average value of ABLH, and Local Time represents the local time. From the average daily variation graphs of each season, the ABLH shows a rising and then falling trend throughout the year, with the height fluctuations gradually decreasing from spring to winter. The maximum value of ABLH in spring (March-May) is about 2.5 km, in summer (June-August) is about 1.3 km, in autumn (September-November) is about 1.8 km, and in winter (December-February) is about 1.5 km. The peak value is reached at 14:00-15:00 for all seasons. The highest value of ABLH appears in spring, which is consistent with the monthly variation graph. The lowest value of ABLH appears in summer, and the difference between the daytime values is smaller than that in other seasons. The possible reason for this is that the LHAASO site is located in the area affected by the Indian summer monsoon. The variation trend of ABLH in the monsoon region is closely related to monsoon activity. During the prevalence of the summer monsoon, with the influence of precipitation, the soil humidity is high, the latent heat is large, and the sensible heat is small, which inhibits the development of the boundary layer. Overall, the ABLH is more stable and the difference is smaller than in other seasons.

\begin{figure}
  \centering
  \includegraphics[width=0.5\textwidth]{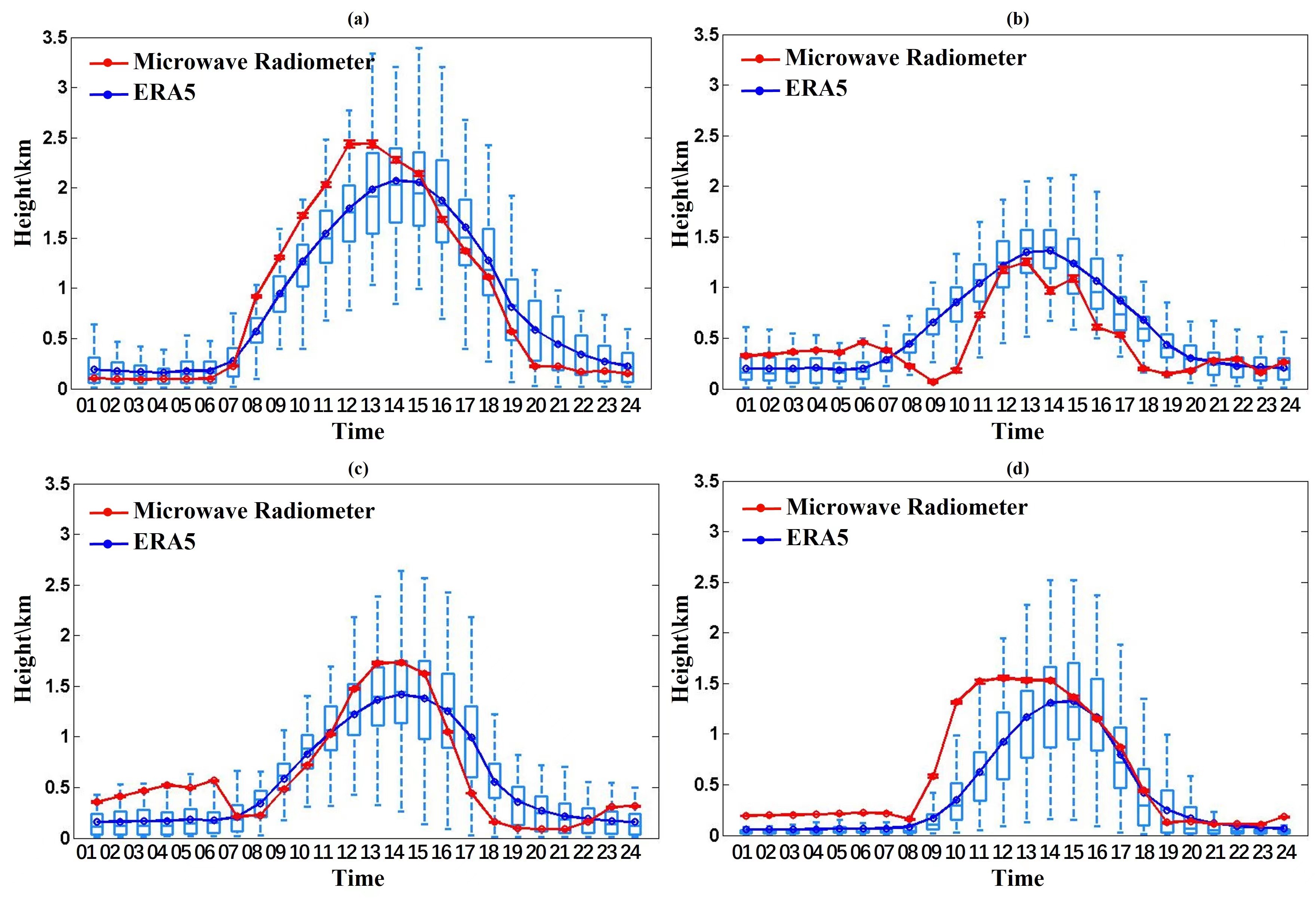}
  \caption{ABLH seasonal mean daily variation (Local Time), (a) Spring, March-May (b) summer, June-August (c) autumn, September-November (d) winter, December-February}
  \label{fig:fig4}
\end{figure}

\section{Conclusion}
Using the temperature profile data observed by the ground-based multi-channel microwave radiometer at LHAASO in Daocheng County, Sichuan Province from February 2021 to November 2021, combined with the potential temperature gradient method, this paper calculated the atmospheric boundary layer height (ABLH) at this site, analyzed the daily, monthly, and seasonal evolution of ABLH, and compared it with the European Centre for Medium-Range Weather Forecasts (ECMWF) ERA5 reanalysis data product, and obtained the following results:

(1) Under clear sky conditions, ABLH exhibits a significant diurnal variation characteristic, and increases continuously with the increase of ground temperature. The ERA5 and microwave radiometer data results have similar trends.

(2) In the monthly variation of ABLH, from spring (March-May), summer (June-August), autumn (September-November) to winter (December-February), ABLH generally shows a decreasing trend. The observed mean value is the highest in April, about 1.2 km, and the lowest in November, with ABLH about 0.6 km. The ERA5 monthly mean heights are slightly lower than the observed values in April and November. The observed ABLH is the highest in spring and the lowest in winter, with summer and autumn in between, which is consistent with the trend shown by ERA5.

(3) In the monthly diurnal variation of ABLH, the daily average values fluctuate slightly. In spring, especially in March and April, the daily average values are significantly higher than those in other months, with mean heights above 1 km. This phenomenon is consistent with the monthly variation trend of microwave radiometer observations.

(4) In the seasonal diurnal variation, ABLH generally shows a downward trend from spring to winter, with an obvious daily variation trend. The ABLH changes greatly during the day, with the highest value usually appearing in the afternoon and being stable at night. The maximum value in spring is significantly higher than that in the other three seasons, which is consistent with the trend of ABLH monthly variation.
Compared with the research on ABLH in other regions of the Qinghai-Tibet Plateau, these conclusions are generally consistent in terms of the diurnal and seasonal variation rules, that is, the ABLH during the day is higher than at night, and the ABLH during the Westerly South Branch circulation period is higher than during the summer monsoon period. The only difference is the numerical value of ABLH. However, compared with the non-monsoon region and the monsoon transition zone (northwest Qinghai-Tibet Plateau), these two regions have the highest ABLH values in summer, while LHAASO is located in the southeast of the Qinghai-Tibet Plateau and is a monsoon region. Influenced by the monsoon, the highest ABLH value occurs in spring.

\section*{Acknowledgments}
This work is supported by the Science and Technology Department of Sichuan Province (grant numbers 2021YFSY0030, 2021YFSY0031), and by National Key R\&D program of China (grant number 2021YFA0718403). We would like to acknowledge the European Centre for Medium-Range Weather Forecasts (ECMWF) for providing ERA5 reanalysis data. The observation data of microwave radiometer observation were obtained from the cooperative observation of the Institute of Plateau Meteorology of Chengdu, China Meteorological Administration in LHAASO.

%Bibliography
\bibliographystyle{unsrt}  
\bibliography{references}

\end{document}